\documentclass[preprint,showpacs]{revtex4}

\usepackage{graphicx}
\usepackage{dcolumn}
\usepackage{amsmath}
\usepackage{xcolor}
\usepackage[colorlinks=true, citecolor=red, urlcolor = red, linkcolor= blue, bookmarks=true]{hyperref}

\begin{document}

\title{Black string surrounded by a static anisotropic quintessence fluid}
\author{Md Sabir Ali $^{a}$} \email{alimd.sabir3@gmail.com, sabir.ali@iitrpr.ac.in}
\author{Fazlay Ahmed$^{a}$} \email{fazleyamuphysics@gmail.com}
\author{Sushant G. Ghosh$^{a,\;b, \;c}$}\email{sghosh2@jmi.ac.in}
\affiliation{$^a$Center for Theoretical Physics,
 Jamia Millia Islamia,  New Delhi 110025
 India}
\affiliation{$^b$ Multidisciplinary Centre for Advanced Research and Studies (MCARS),\\ Jamia Millia Islamia, New Delhi 110025, India}
\affiliation{$^c$Astrophysics and Cosmology Research Unit,
 School of Mathematics, Statistics and Computer Science,
 University of KwaZulu-Natal, Private Bag X54001,
 Durban 4000, South Africa}

\begin{abstract}
A black string is a solution of the four-dimensional general relativity in cylindrical symmetric anti-de Sitter (AdS) spacetime pioneered by Lemos \cite{Lemos:1994xp}. We obtain an exact rotating black string surrounded by quintessence matter in anti-de Sitter (AdS) spacetime and analyze its properties. The solution has an additional parameter $N_q$ due to background quintessence matter and it also depends on quintessence state parameter $w_q$. Our rotating solution encompasses the Lemos black string \cite{Lemos:1994xp}, which can be recovered in the absence of background matter ($N_q=0$). The special case of rotating charged black string ($w_q=1/3$) has both Cauchy and event horizons. We find that thermodynamic quantities mass, temperature, entropy and heat capacity get corrected due to quintessence background. The counterterm method is utilized to calculate the associated conserved quantities to conclude that the first law of thermodynamics is satisfied. The cloud of strings background ($w_q= -1/3$) has been included as a special case. 
\end{abstract}

\maketitle

\section{Introduction}
The cosmic censorship conjecture \cite{Penrose:1969pc} states that singularities formed from gravitational collapse with regular initial surfaces must always be hidden behind an event horizon, which means no light rays breathe out of singularity, i.e., singularity is never naked. Despite almost more than 50 years of effort, it still remains an open problem (for reviews and references, see \cite{Joshibook}). On the other hand, several investigations of the gravitational collapse, particularly in cylindrical symmetry are motivated by Thorne’s hoop conjecture \cite{Thorne1972}. The gravitational collapse will lead to a black hole if a mass $M$ is compacted into a region whose circumference in all direction $C \leq 4 \pi M$ \cite{Penrose:1969pc}. Thus, the gravitational collapse of planar or cylindrical matter will not form a black hole (black plane or black string) \cite{Penrose:1969pc}. Thus, a naked singularity may form in planar or cylindrical relativistic collapse if it results into $C > 4 \pi M$ in some direction violating cosmic censorship conjecture, but not the hoop conjecture \cite{Ghosh:2000ud}. However, the hoop conjecture was formulated for spacetime with a zero cosmological term and it is possible that in the presence of a cosmological term situation may change drastically. Indeed, Lemos \cite{Lemos:1998iy} has investigated the gravitational collapse of a planar or cylindrical matter distribution (null fluid) in an anti-de Sitter (AdS) space-time to show that black plane or black string form rather than naked singularity, violating in this way the hoop conjecture but not cosmic censorship conjecture. Here, as in the BTZ black holes, negative cosmological constant plays a crucial role as in the final fate of the gravitational collapse. The Black string may be regarded as counterpart of spherically symmetric Schwarzschild black holes in cylindrical symmetric AdS spacetimes. Lemos \cite{Lemos:1998iy,Lemos:1994fn,Lemos:1994xp} pioneered research on the black string and soon its charged and rotating \cite{Lemos:1995cm} counterpart were also discovered by him. These black string solutions in four dimensions are asymptotically anti-de Sitter in both the transverse and the string directions. Horowitz and Strominger \cite{Horowitz:1991cd} has shown that black string solution in the transverse direction of asymptotically flat spacetime does not exist. Subsequently, there has been intense activities in the investigation of the black string \cite{Brihaye:2005tx,Brihaye:2006hsa,Aliev:2011ir,Kleihaus:2012qz,Tannukij:2017jtn,Cisterna:2017qrb}, and more recently on rotating  black strings \cite{Sheykhi:2008rk,Sheykhi:2013yga,Hendi:2010kv,Hendi:2013mka,Singh:2017bwj}. 

Recent cosmological observations indicate that besides the baryonic and dark matter, there exists a dark energy candidate, which is responsible for the accelerated expansion of the observed Universe, whose origin and nature is still unknown \cite{Ade:2013sjv}. One possible candidate to explain nature of the dark energy is the quintessence field that is characterized by the equation of state $p_q=w_q\rho_q$, where $p_q$ and $\rho_q$ are, respectively, the pressure and energy density of quintessence field, and $w_q$ is the equation of state parameter lies in the interval $-1<w_q<-1/3$. The first black hole solution in the quintessence background was obtained by Kiselev \cite{Kiselev:2002dx}, and its charged solution was obtained in a ref. \cite{Azreg-Ainou:2014lua}. Further, Kiselev's solution has been extended to higher dimensions \cite{Chen:2008ra}, and later in five-dimensional Einstein-Gauss-Bonnet gravity \cite{Ghosh:2016ddh}, and Lovelock gravity \cite{Ghosh:2017cuq}. The rotating counterpart of Kiselev solution has been also obtained \cite{Ghosh:2015ovj,Toshmatov:2015npp}. The thermodynamics of black holes in the quintessence background has been also investigated \cite{Wei:2011za,AzregAinou:2012hy,Tharanath:2013jt,Ghaderi:2016dpi}, and so is the quasinormal modes \cite{Zhang:2007nu,Saleh:2011zz,Tharanath:2014uaa}.

The aim of this paper is to construct an exact rotating black string surrounded by quintessence matter in Anti-de Sitter (AdS) spacetime. We discuss the horizon structure and also examine the thermodynamic aspects of the solutions to discuss the effect of surrounding quintessence matter. We also analyze the thermal stability of the black hole solutions by performing the study of heat capacity. It may be mentioned that the investigation of the black hole/string solutions in AdS space has primary relevance for the AdS/CFT correspondence \cite{Maldacena:1997re,Witten:1998qj,Susskind:1994vu,Aharony:1999ti}. 
The structure of this paper is as follows. The Sec.~\ref{solution} is devoted to find the solution of rotating black string surrounded by a quintessence matter. We use the counterterm method to calculate the conserved quantities associated with the solution in Sec.~\ref{massangular}. The thermodynamics of the black string is the subject of Sec.~\ref{thermodynamics}, and in this section, we also include a special case of the cloud of strings background. We complete the paper by concluding remarks in Sec.~\ref{conclusion}.

\section{Rotating black string solution}
\label{solution}
The effective action, in the presence of a cosmological constant, reads \cite{Hendi:2010kv,Lemos:1994xp}
\begin{equation}
\label{action}
S_G=-\frac{1}{16\pi}\int_{\mathcal{M}}d^{4}x\sqrt{-g}(R-2\Lambda)-\frac{1}{8\pi}\int_{\partial\mathcal{M}}d^{3}x\sqrt{-\beta}\vartheta+ S_{f},
\end{equation}
where $R$ is Ricci scalar, $\Lambda=-3 l^{-2}$ is a cosmological constant, $\beta$ is a trace of the induced metric, $\vartheta$ is extrinsic curvature defined on the boundary $\partial \mathcal{M}$, and $S_{f}$ is the action related to the matter field. Varying action (\ref{action}) with respect to $g_{\mu\nu}$, we obtain the following field equations
\begin{equation}
\label{fe}
R_{\mu\nu}-\frac{1}{2}g_{\mu\nu} (R-2\Lambda)=T_{\mu\nu},
\end{equation}
where $T_{\mu \nu}$ is the energy-momentum tensor, which for the quintessence matter \cite{Kiselev:2002dx} is given by
\begin{eqnarray}
\label{fe2}
T^t{_t}&=&T^r{_r}=-\rho_{q},\nonumber\\
T^\phi{_\phi}&=&T^z{_z}=\frac{1}{2}\rho_q\left(3w_q+1\right).
\end{eqnarray}
Thus, the quintessence matter obeys the equation of state
\begin{eqnarray}\label{eos}
p=\frac{1}{2}\rho_q (3 w_q+1),
\end{eqnarray}
where $\rho_{q}$ is the energy density and $w_q$ is the equation of state parameter for the quintessence field. Obviously, $T^t_t=-\rho_q=T^r_r=p_r$, $T^{\phi}_{\phi}=p_{\phi}$, and $T^z_z=p_z$, and the fluid is anisotropic as $p_r\neq p_{\phi}$. Our main interest is to obtain an exact rotating black string solution surrounded by the quintessence matter. We begin with a general static black string metric in ($\overline{t},r,\overline{\phi},z$) coordinates \cite{Lemos:1994xp}
\begin{eqnarray}
\label{metricquitessence}
ds^{2}&=&-f(r)d \overline{t}^2+\frac{dr^2}{f(r)} +r^2 d\overline{\phi}^2+ \frac{r^2}{l^2} dz^2.
\end{eqnarray}
The 2-dimensional space $t=r=$ constant has a topology $R\times S^{1}$ such that $-\infty<\overline{t}<\infty$, $0\leq r<\infty$, $0\leq \overline{\phi}<2\pi$, $-\infty<z<\infty$. 
Using Eqs.~(\ref{fe}), (\ref{fe2}), and (\ref{metricquitessence}), we obtain the following Einstein's field equations
\begin{eqnarray}
\label{fe1}
&&\frac{f^\prime(r)}{r}+\frac{f(r)}{r^2}-3l^{-2}=-\rho_q, \nonumber\\
&& \frac{f^{\prime\prime}(r)}{2}+\frac{f^\prime (r)}{r}-3l^{-2}=p.
\end{eqnarray}
On using equation of state (\ref{eos}), we get
\begin{eqnarray}
\label{master}
r^2 f^{\prime\prime}+3 r f^\prime(r) \left(w_q+1\right)+f(r) \left(3w_q+1\right)-9r^2l^{-2}\left(w_q+1\right)=0.
\end{eqnarray} 
Eq.~(\ref{master}) can be integrated to get
\begin{eqnarray}\label{fr}
f(r)=\frac{r^2}{l^2}-\frac{2m}{r}+\frac{N_q}{r^{3w_q+1}},
\end{eqnarray}
where $m$ and $N_q$ are integration constants, and $l$ is the curvature radius of AdS. 
Using Eqs.~(\ref{fe1}) and (\ref{fr}), the density parameter $\rho_q$ of the quintessence field reads
\begin{eqnarray}
\rho_q=\frac{N_q}{2}\frac{3w_q}{r^{3(w_q+1)}},
\end{eqnarray}
which is always positive. Since for quintessence $w_q\leq0$, then $N_q$ is restricted to be negative. \\
Next, we discuss the validity of the black string solution in the framework of tetrad formalism. The basis vectors of the local observer and the basis vectors of the metric (\ref{metricquitessence}) are related by
\begin{eqnarray}
e_{(a)}=e^{\mu}_{(a)}\partial_\mu,\qquad g^{\mu\nu}=\eta^{(a)(b)}e^{\mu}_{(a)}e^{\nu}_{(b)},
\end{eqnarray}
where $\eta^{(a)(b)}=\text{diag}(-1,1,1,1)$. In the coordinates $(\overline{t},r,\overline{\phi},z)$ to describe static solutions (\ref{metricquitessence}) with metricfunction $f(r)$ given in Eq.~(\ref{fr}), we introduce the basis vectors
\begin{eqnarray}
\label{tetrads}
e^{(a)}_\mu=\text{diag}(\sqrt{\frac{r^2}{l^2}-\frac{2m}{r}+\frac{N_q}{r^{3\omega_q+1}}},\frac{1}{\sqrt{\frac{r^2}{l^2}-\frac{2m}{r}+\frac{N_q}{r^{3\omega_q+1}}}},r,\frac{r}{l})
\end{eqnarray}
The components of the energy-momentum tensor in the orthonormal basis are written as
\begin{eqnarray}
\label{emt1}
T^{(a)(b)}=e^{(a)}_\mu e^{(b)}_\nu G^{\mu\nu},
\end{eqnarray}
which equivalently are written as
\begin{eqnarray}
\label{emt2}
G_{(a)(b)}=e^\mu_{(a)}e^\nu_{(b)}T_{\mu\nu}.
\end{eqnarray}
The orthonormal tetrads together with the enermy-momentum tensor are evaluated to be
\begin{eqnarray}
G_{(t)(t)}=-G_{(r)(r)}=-3/l^2+\frac{3N_q\omega_q}{r^{3\left(\omega_q+1\right)}},\qquad G_{(\phi)(\phi)}=G_{(z)(z)}=3/l^2+\frac{3N_q\omega_q(3\omega_q+1)}{2r^{3\left(\omega_q+1\right)}}.
\end{eqnarray}
Hence we can write
\begin{eqnarray}
\label{p1}
\rho=-p_r=-3/8\pi l^2+\frac{3N_q\omega_q}{8\pi r^{3\left(\omega_q+1\right)}},\qquad p_\phi=p_z=3/8\pi l^2+\frac{3N_q\omega_q(3\omega_q+1)}{16\pi r^{3\left(\omega_q+1\right)}}.
\end{eqnarray} 
Therefore, we can see that the resulting fluid in neither a perfect nor it is isotropic. Following \cite{Visser:2019brz}, the average pressure of the resulting fluid is evaluated to be
\begin{eqnarray}
\label{avep}
\overline{p}=\frac{p_r+p_\phi+p_z}{3}=3/8\pi l^2+\frac{3N_q\omega_q}{8\pi r^{3\left(\omega_q+1\right)}},\qquad \frac{\overline{p}}{\rho}=\frac{1/l^2+\frac{N_q\omega_q^2}{r^{3(\omega_q+1)}}}{-1/l^2+\frac{N_q\omega_q^2}{r^{3(\omega_q+1)}}}.
\end{eqnarray}
Such average pressure is also not resulted into the perfect fluid equation, instead it has dependence on the radial coordinate because of the presence of the $1/l^2$-term. If we put $1/l^2=0$, we are landing with situation when black string solution does not exist as it demands the presence of the cosmological constant term. Further to be more specific about the pressure, we calculate the pressure ratio and pressure anisotropy as follows
\begin{eqnarray}
\frac{p_\phi}{p_r}=\frac{l^2+\frac{N_q\omega_q(3\omega_q+1)}{2 r^{3\left(\omega_q+1\right)}}}{l^2-\frac{N_q\omega_q}{r^{\left(\omega_q+1\right)}}},\qquad \delta=\frac{p_r-p_\phi}{\overline{p}}=\frac{3}{2}\frac{N_q\omega_q\left(\omega_q+1\right)}{r^{3(\omega_q+1)}/l^2+N_q\omega_q}.
\end{eqnarray}
We see that the ratio $p_t/p_r$ and the relative pressure anisotropy $\delta$ are position dependent and therefore the identification of the black string as perfect and isotropic solution is lost. Hence, the black string solution is imperfect and anisotropic and the demand of the article \cite{Visser:2019brz} that ``\textit{The Kiselev solution is neither perfect fluid, nor it is quintessence"} is identically hold good.
Thus, the metric~(\ref{metricquitessence}) with (\ref{fr}) represents an exact non-rotating static black string surrounded by an anisotropic quintessence matter.
The Kretschmann scalar for the metric (\ref{metricquitessence}) with (\ref{fr}) is
\begin{eqnarray}\label{krtch}
R_{\mu\nu\alpha\beta}R^{\mu\nu\alpha\beta}&=&\frac{24 N_q}{l^4}+\frac{48 m^2 N_q}{r^6} -\frac{24 (3 w_q+2)(w_q+1)m N_q}{r^{3 w_q+6}}-\frac{w_q(3w_q-1)N_q}{l^4 r^{3w_q+3}} \nonumber\\ &&+\frac{(27 w_q^4+54 w_q^3+51 w_q^2+20 w_q+4)N_q}{r^{6 w_q+6}}.
\end{eqnarray}
It is seen that for $N_q\neq0$, $m\neq0$, the Kretschmann scalar diverges at $r=0$. 
The metric (\ref{metricquitessence}) admits three Killing vectors, $\partial/\partial t$, $\partial/\partial z$, and $\partial/\partial \phi$, respectively, correspond to time-translation along $t$-axis, translation symmetry along $z$-axis, and rotational symmetry around the $\phi$-axis.
The rotating counterpart of metric~(\ref{metricquitessence}) can be obtained by using the transformations \cite{Lemos:1998iy}
\begin{eqnarray}
\label{trf}
t=\Xi \overline{t}-a\overline{\phi},\quad \phi=\Xi \overline{\phi}-\frac{a}{l^2}\overline{t},
\end{eqnarray}
where $\Xi=\sqrt{1+a^2/l^2}$ and $a$ is the rotation parameter. Substituting (\ref{trf}) and (\ref{fr}) into (\ref{metricquitessence}), we obtain
\begin{eqnarray}
\label{rotBSQ}
ds^{2}&=&-\left(\frac{r^2}{l^2}-\frac{2m}{r}+\frac{N_q}{r^{3w_q+1}}\right) \left(\Xi dt-a d\phi\right)^2+\frac{dr^2}{\left(\frac{r^2}{l^2}-\frac{2m}{r}+\frac{N_q}{r^{3w_q+1}}\right)} \nonumber\\ &&+\frac{r^2}{l^4}\left(adt-\Xi l^2d\phi\right)^2 +\frac{r^2}{l^2} dz^2.
\end{eqnarray}
The metric~(\ref{rotBSQ}) represents a rotating stationary black string surrounded by quintessence matter, which solves field Eqs.~(\ref{fe}). The metric encompasses vacuum black string \cite{Lemos:1998iy} as a special case when $N_q=0$. Also, for $w_q=-1$, it again goes over to vacuum black string but with an effective cosmological constant  
\begin{eqnarray}
1/l_{eff}^2=\left(1/l^2+N_q\right).
\end{eqnarray} 
The charged black string \cite{Lemos:1995cm,Hendi:2010kv} can be obtained by choosing $w_q=1/3$, and the black string surrounded by cloud of strings for $w_q=-1/3$. The metric~(\ref{rotBSQ}) has many properties similar to the Kerr metric on equatorial plane \cite{Lemos:1995cm}. For $N_q=-1/l^2$, the cosmological constant term disappears from the solution and we get no black string. However, for $N_q>0$ one could expect an anti-quintessence effect with the negative energy density $(\rho_q<0)$. Subsequently a bare cosmological constant could be built up if one set $N_q=1/l^2$ for the state parameter $w_q=-1$. 

  \begin{figure}[t]
\includegraphics[scale=0.5]{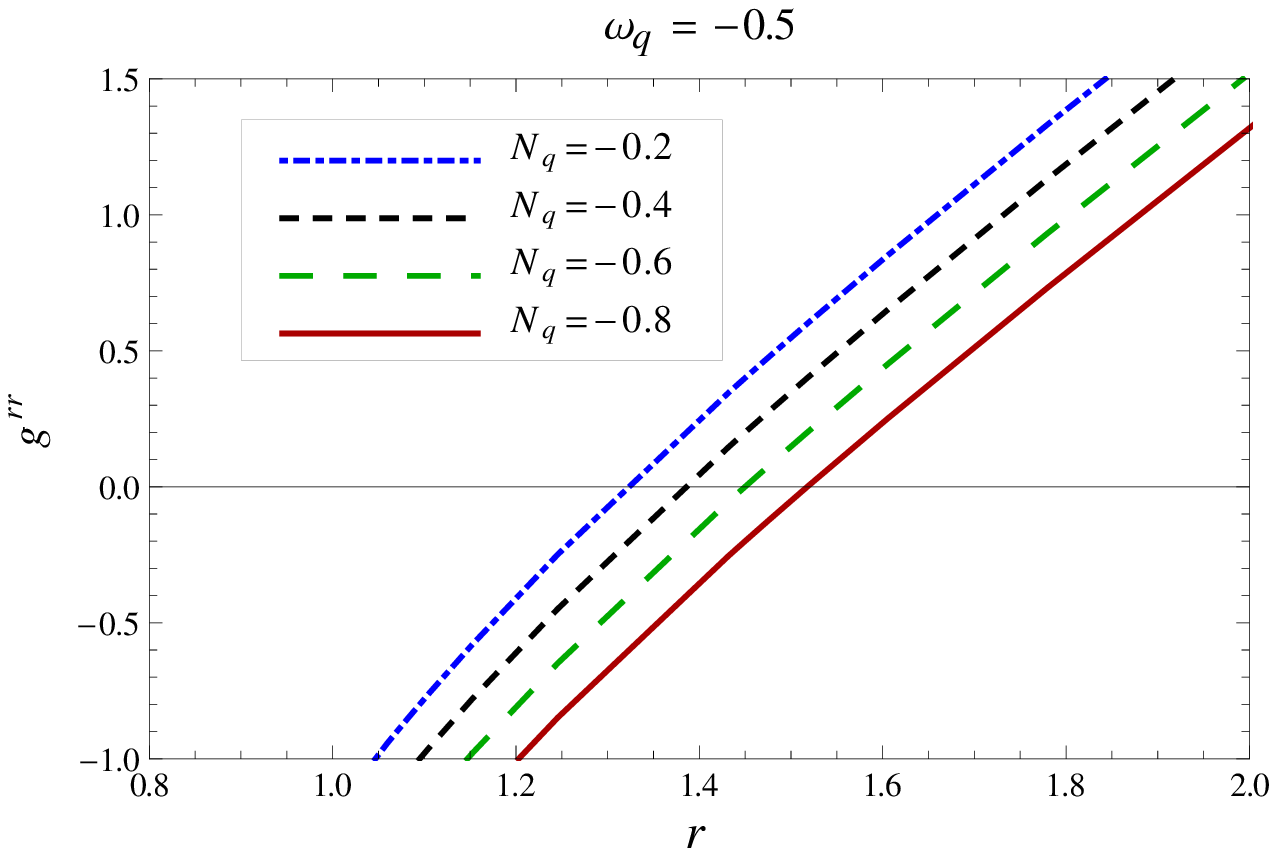}
\includegraphics[scale=0.5]{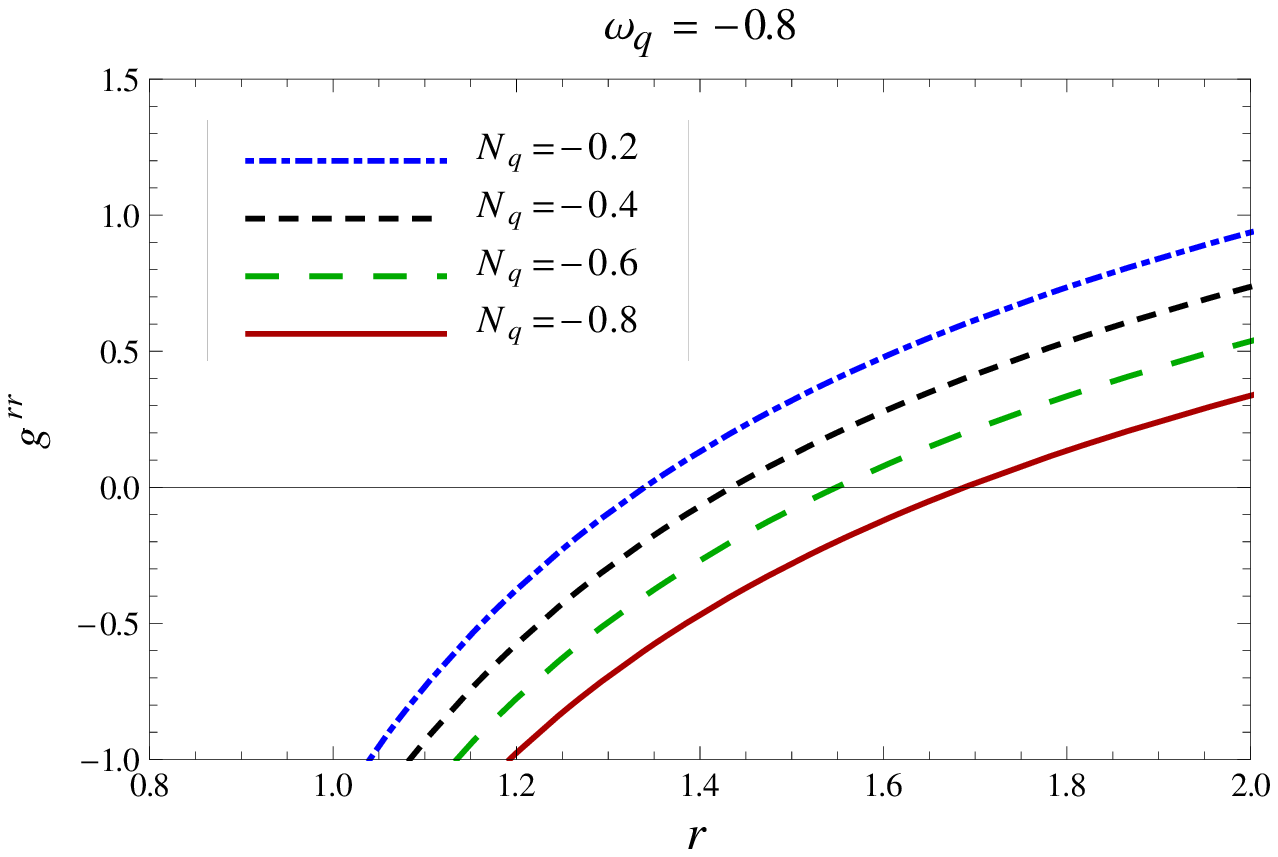}
 \caption{Plots showing the horizons of rotating black string sorrounded by quintessence for various values of $N_q$ in units of $l^{3w_q+1}$. Here the radial coordinate $r$ is written in units of $l$. \label{fig1}}
\end{figure}
\begin{figure}[t]
\includegraphics[scale=0.5]{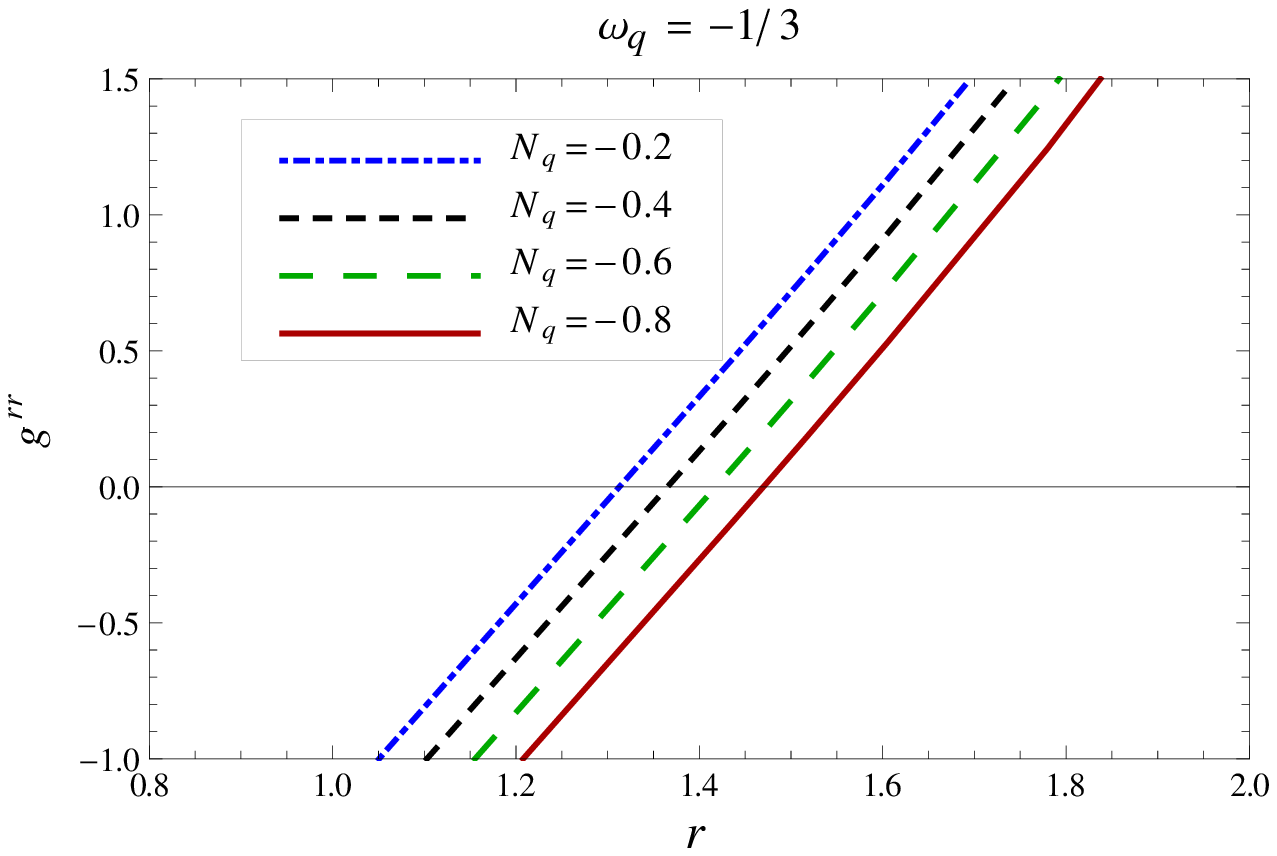}
\includegraphics[scale=0.5]{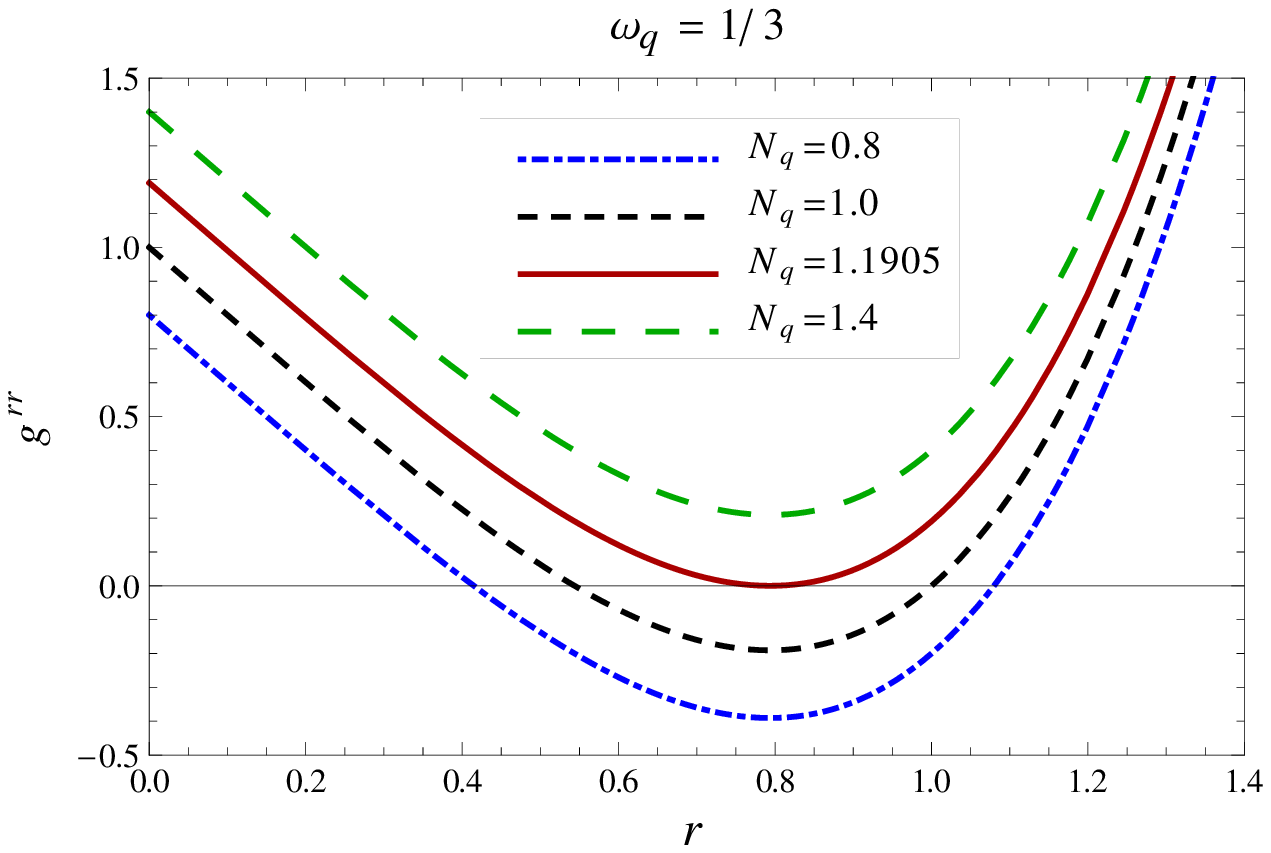}
 \caption{Plots showing the horizons of for string cloud background $(left)$, and charged case $(right)$ for various values of $N_q$ in units of $l^{3w_q+1}$. Here the radial coordinate $r$ is written in units of $l$. \label{fig5}}
\end{figure}
The horizon of the solution~(\ref{rotBSQ}) can be obtained by solving $g^{rr}=0$, i.e.,
\begin{eqnarray}\label{horizonqnt}
r^{3 w_q+3}-2 m l^2 r^{3 w_q}+l^2 N_q=0. 
\end{eqnarray}
Solving Eq.~(\ref{horizonqnt}) numerically for appropriate values of parameters is shown in Fig.~\ref{fig1}. One can see that the radius of horizon decreases as the value of parameter $N_q$ increases, and for $w_q$, horizon radius increases with an increase in $w_q$.
The horizon for the cloud of strings ($w_q=-1/3$) background is
\begin{eqnarray}
r_+ = \frac{ \left(27 l^2 m+\sqrt{729 l^4 m^2+27 l^6 N_q^3}\right)^{2/3} -3 l^2 N_q}{3 \left(27 l^2 m+\sqrt{729 l^4 m^2+27 l^6 N_q^3}\right)^{1/3}}.
\end{eqnarray}
It is clear that for negative values of $w_q$ there exist a single horizon (cf. Figs.~\ref{fig1}, \ref{fig5}). Thus, the horizon structure of rotating black string in quintessence background is different from the Kerr black hole. However, for $w_q=1/3$ the charged black string solution admitting two horizons  
\begin{eqnarray}
\label{horQ}
r_{\pm}=\frac{12l^2N_q+y^{2/3}}{3y^{1/3}}\pm\sqrt{\frac{4ml^2}{\sqrt{\frac{12l^2N_q+y^{2/3}}{3y^{1/3}}}}-\frac{12l^2N_q+y^{2/3}}{3y^{1/3}}},
\end{eqnarray}
where $y=54m^2l^4+\sqrt{1296m^4l^8-1728l^6N_q^3}$. The $``+"$ sign in Eq.~(\ref{horQ}) denotes the event horizon $r_+$ whereas the $``-"$ sign represents the Cauchy horizon $r_c$, respectively. The horizons of the charged rotating black string are depicted in Fig.~{\ref{fig5}}. We have two horizons when $r_+>r_-$, the degenerate horizons for $r_+=r_-$, and no black string when the Eq.~(\ref{horQ}) corresponds to no real root. Thus, only the charged black strings in (A)dS universe has two horizons viz.,  Cauchy horizon and event horizon. However, in the black string surrounded by quintessence backgrounds, the Cauchy horizon disappears and it has just a single horizon. Thus, Cauchy horizons are unstable. 

Next, we comment on the orthonormal basis of the rotating quintessence black string in which the energy-momentum tensor is diagonal. The orthonormal tetrads are written as 

\begin{eqnarray}
   e^{(a)}_\mu &=&
  \left( {\begin{array}{cccc}
   \sqrt{f(r)}\Xi & 0 & -\sqrt{f(r)}a & 0\\
   0 & \frac{1}{\sqrt{f(r)}} & 0 & 0\\
    \frac{a\;r}{l^2} & 0 & -r\Xi & 0 \\
0 & 0 & 0 & \frac{r}{l} \\
  \end{array} } \right).
\end{eqnarray}
The components of the energy–momentum tensor in the orthonormal frame read
\begin{eqnarray}
T^{(a)(b)}=e^{(a)}_\mu e^{(a)}_\nu G^{\mu\nu}.
\end{eqnarray}
Fortunately, while calculating for the rotating quintessence black string we ends with same calculations for density $\rho$ and pressure $(p_r,\;p_\phi,\;p_z)$ as in the nonrotating case. 
\section{Conserved quantities associated with black string}\label{massangular}
Here, we calculate the conserved quantities associated with the rotating black string by counterterm approach \cite{Balasubramanian:1999re}. The gravitational action for four-dimensional asymptotically AdS spacetime $\mathcal{M}$ in the presence of some field \cite{Hendi:2010kv} reads
\begin{eqnarray}
\label{actiontot}
S_{tot}= S_G-\frac{1}{4\pi} \int_{\partial \mathcal{M}} d^3 x \sqrt{-\beta} \left(-\frac{1}{l}\right),
\end{eqnarray}
Following the counter formalism and varying the action (\ref{actiontot}), we can write the divergence free boundary energy-momentum tensor as
\begin{equation}
T^{ab}=\frac{1}{8 \pi} (\vartheta^{ab}-(\vartheta+2 l^{-2})\beta^{ab}).
\end{equation}
 To calculate the conserved charges of the spacetime, one should choose a spacetime surface $\mathcal{B}$ in $\partial \mathcal{M}$ with metric $\sigma_{ij}$ and write the metric in Arnowitt-Deser-Misner (ADM) \cite{Hendi:2010kv} form
\begin{equation}\label{gammaab}
\beta_{ab} dx^{a} dx^{b} = -N^2 dt^2 +\sigma_{ij}(d\phi^i+V^i dt)(d\phi^j+V^j dt).
\end{equation}
Here the coordinates $\phi^i$ are the angular variables at the hypersurface of constant $r$ around the origin. Here $N$  and $V^i$ are the lapse and shift functions, respectively. We define a Killing vector field $\xi^{\mu}$ on the boundary, the conserved quantities related to the energy-momentum tensor of Eq.~(\ref{gammaab}) reads
\begin{equation}
Q_{\xi}= \int_{\mathcal{B}} d^2 x \sqrt{\sigma} T_{ab} n^{a} \xi^{b},
\end{equation}
where $\sigma$ is the determinant of the metric $\sigma_{ij}$, and $n^{a}$ is the associated unit normal vector on the boundary $\mathcal{B}$. The boundary includes both the time-like Killing vector $\xi^{\mu}_{(t)}$ and a rotational Killing vector $\xi^{\mu}_{(\phi)}$ fields. The conserved quantities associated with these two conserved fields are
\begin{eqnarray}
M=\int_{\mathcal{B}} d^2x \sqrt{\sigma} T_{ab} n^a \xi^b_{t}, \nonumber\\
J=\int_{\mathcal{B}} d^2x \sqrt{\sigma} T_{ab} n^a \xi^b_{\phi}.
\end{eqnarray}
One can find the quasilocal mass and the angular momentum per unit length of the string when the boundary $\mathcal{B}$ goes to infinity as
\begin{eqnarray}
\label{conserved}
M &=& \frac{1}{16 \pi l} (3 \Xi^2-1)m, \nonumber\\ 
J &=& \frac{3}{16 \pi l} \Xi m a.
\end{eqnarray}
where $m$ is the mass of the black string. For $a=0$, the angular momentum vanishes from Eq.~\ref{conserved}.

\section{Thermodynamics of rotating black string}\label{thermodynamics}
Next, we calculate the thermodynamic quantities of the rotating black string. It is well known that the universal area law equally is applicable in all black objects in Einstein gravity \cite{Bekenstein:1973ur,Bekenstein:1983iq,Gibbons:1977mu,Hawking:1998ct,Hawking:1998jf}. Therefore the entropy per unit length of the black string reads
\begin{equation}
S_+=\frac{\pi\Xi r_{+}^2}{4 l}.
\end{equation}
The entropy of the black string surrounded by quintessence matter still obeys the area law which states that the entropy of the black hole is a fourth of the event horizon area
\begin{figure}[t]
\includegraphics[scale=0.5]{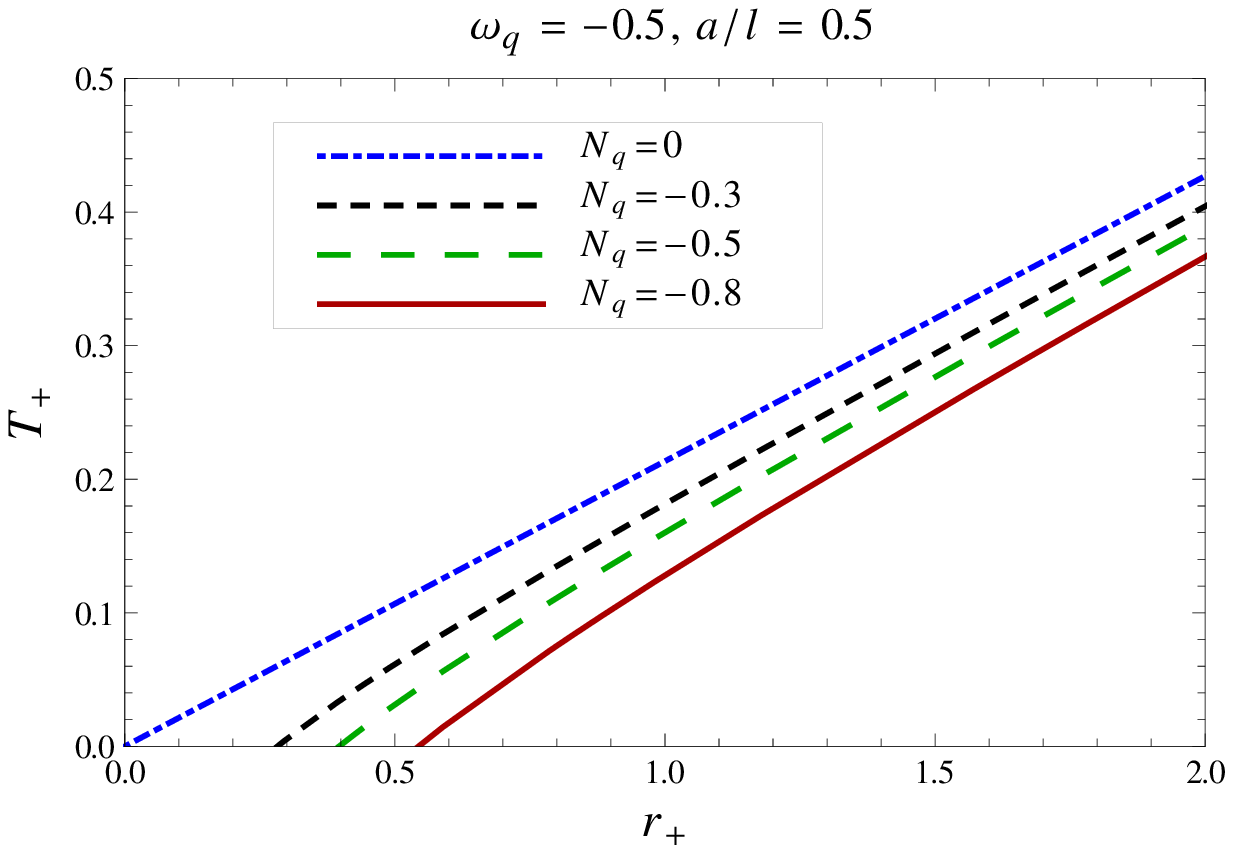}
\includegraphics[scale=0.5]{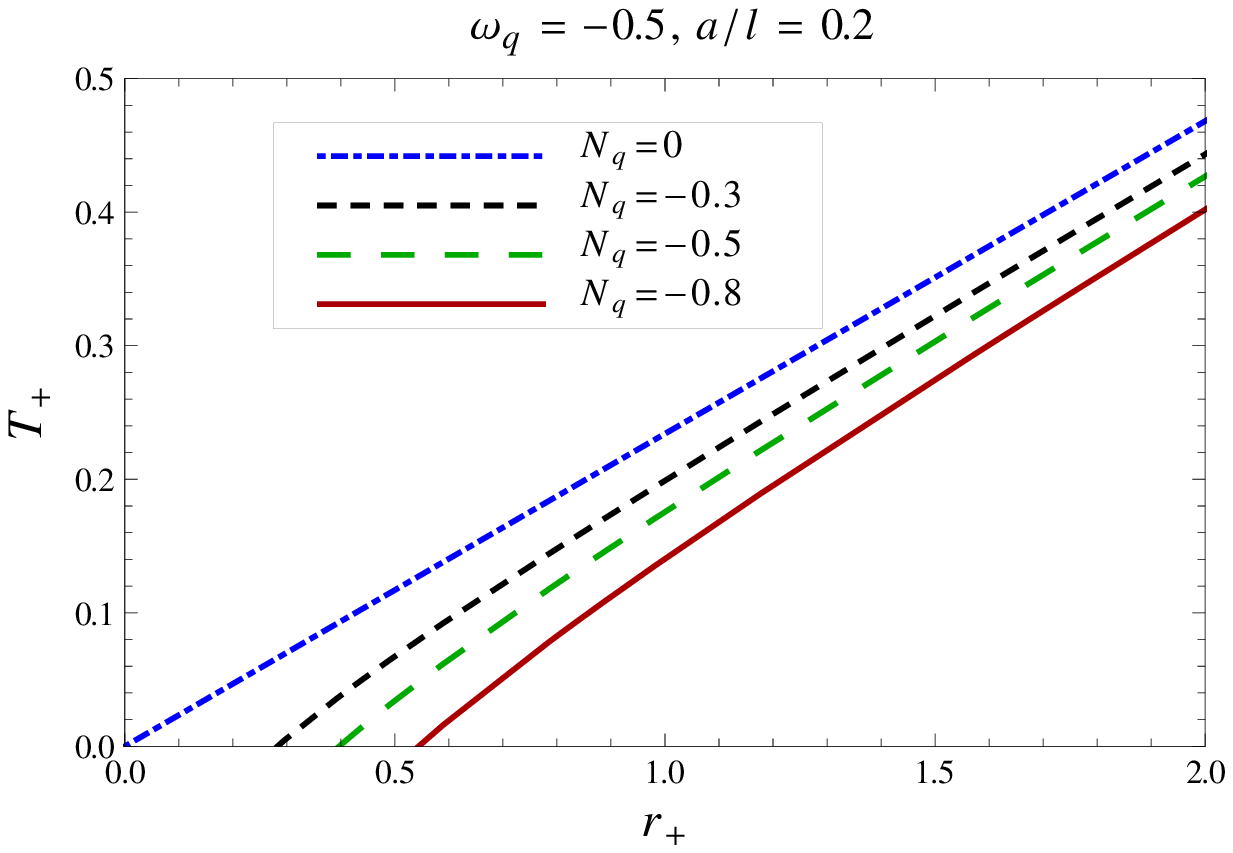}\\
\includegraphics[scale=0.5]{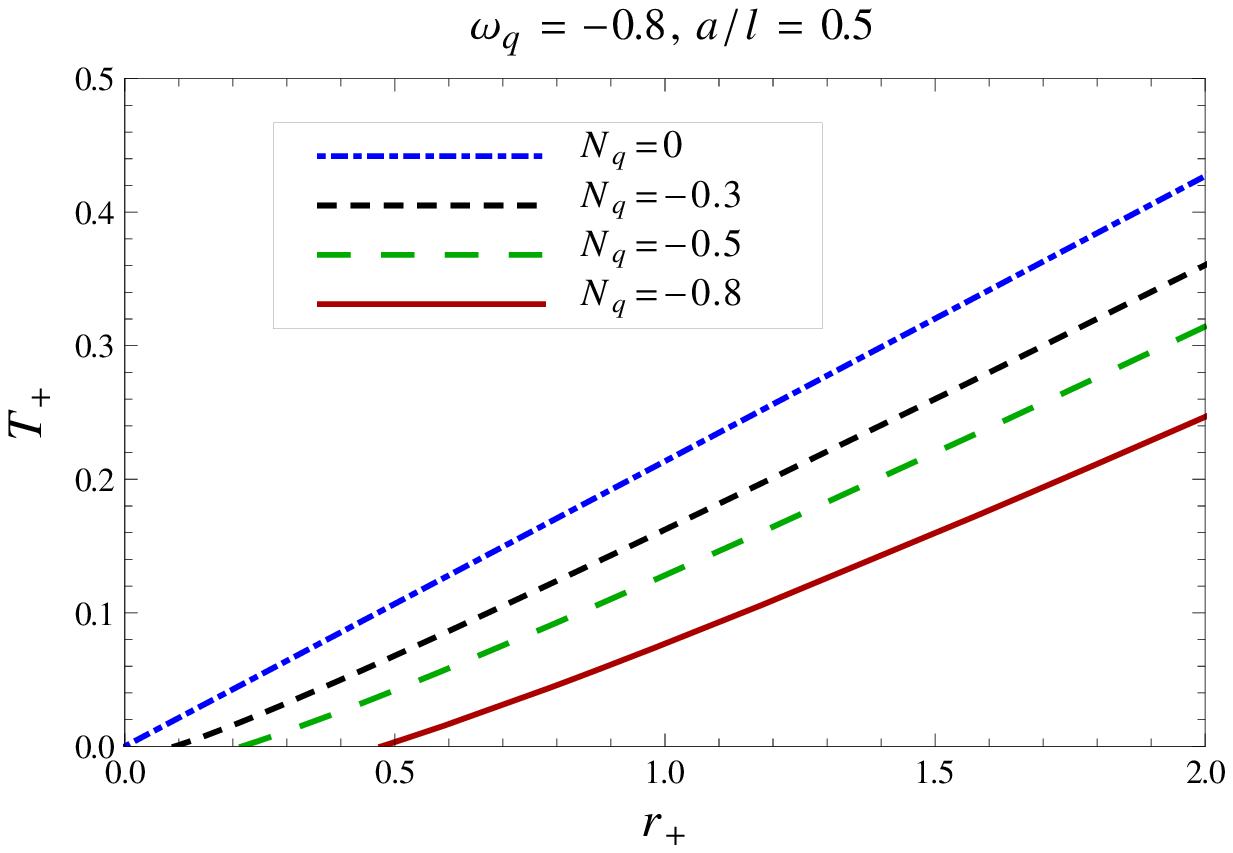}
\includegraphics[scale=0.5]{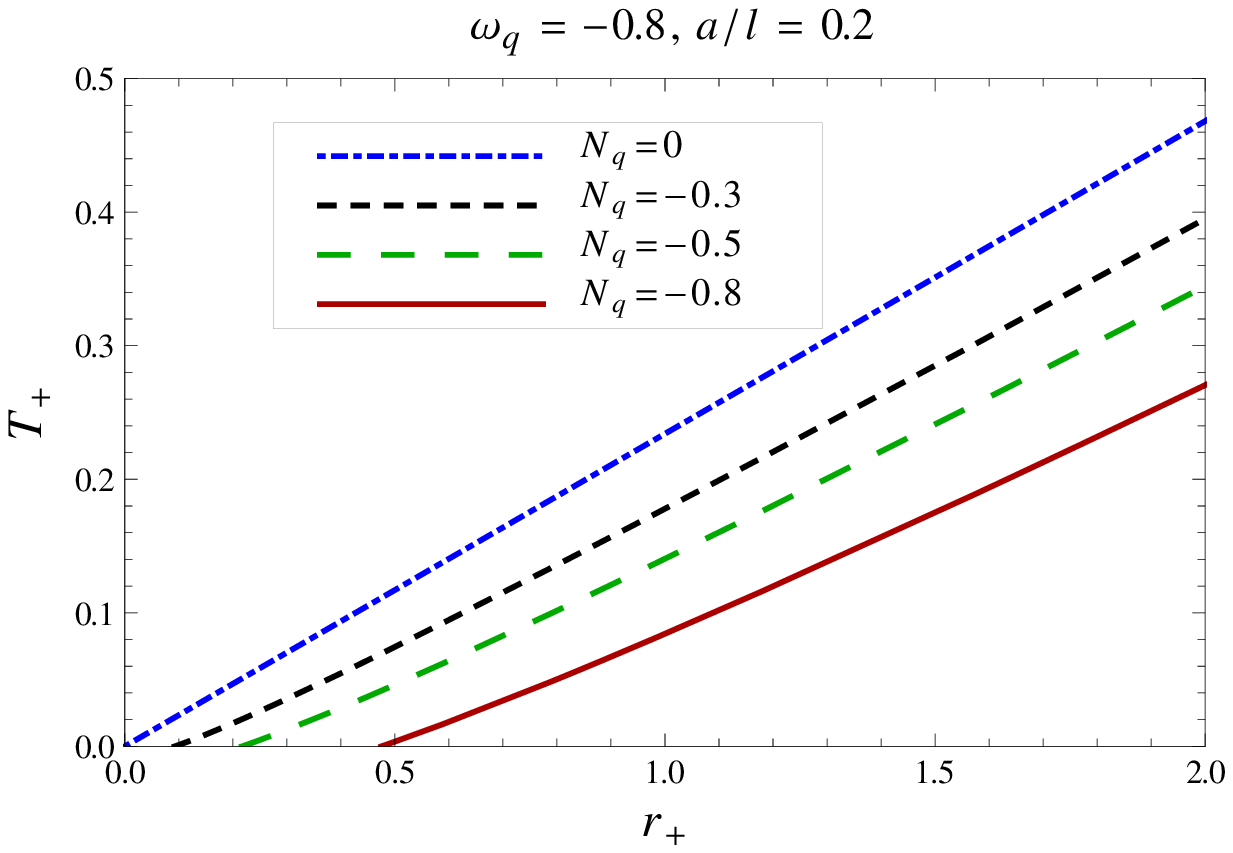}
 \caption{Plots showing the behaviour of temperature $T_+$ with horizon radius $r_+$ for rotating black string sorrounded by quintessence for various values of $N_q$ in units of $l^{3w_q+1}$. Here the horizon radius $r_+$ is written in units of $l$ and temperature is written in units of $1/l$. \label{fig2}}
\end{figure}
The rotating black string with axial symmetry admits two Killing vectors, namely, $\xi^\mu_{(t)}$ corresponding to the time-translational symmetry along $t$-axis and $\xi^\mu_{(\phi)}$ corresponding to the rotational symmetry about $\phi$-axis. These two Killing vectors generate a Killing field $\chi^\mu$=$\xi^\mu_{(t)}+w_+\xi^\mu_{(\phi)}$, where $w_+$ is the angular velocity at the event horizon. Using the analytic continuation method, we obtain the angular velocity and related temperature of the black string at the event horizon. The Euclideanizing the Lorentzian metric by the transformation $t \rightarrow i \tau$, and $a \rightarrow ia$, in such a way that for regular behavior of the coordinates at $r=r_{+}$ demands that $\tau \rightarrow \tau \beta_{+}$ and $\phi \rightarrow \phi+iw_+\beta_{+}$, where $\beta_{+}$ is the inverse Hawking temperature. For rotating black string the Hawking temperature is written as
\begin{eqnarray}
\label{temp}
T_+=\beta_+^{-1}=\frac{3}{4 \pi r_+ \Xi}\left(\frac{r_+^2}{l^2}-\frac{N_q w_q}{r_+^{3 w_q+1}}\right).
\end{eqnarray}
It can be seen that the temperature (\ref{temp}) is always positive when $-1<w_q<-1/3$. We plot the Hawking temperature in terms of horizon radius in Fig.~\ref{fig2} for different values of $a$, $w_q$ and $N_q$. For a fixed value of the state parameter $w_q$, the temperature shifts towards the higher values when the value of quintessence parameter $N_q$ increases. When $N_q=0$, the temperature becomes
\begin{eqnarray}
T_+=\frac{3r_+}{4 \pi\Xi l^2}.
\end{eqnarray}
Thus, the temperature behaves like a straight line passing through the origin (c.f.~Fig.~\ref{fig2}).
The angular velocity $\Omega_+$ is calculated to be 
\begin{equation}
\Omega_+=\frac{a}{\Xi l^2}.  
\end{equation}
We are now in a position to check the first law of black hole thermodynamics. The Smarr type relation for a rotating black string in the presence of quintessence can be written as
\begin{equation}
M= 2 T_+ S_+ + 2 \Omega_+ J_+ -(3 w_q +1) N_{q} \Theta_{q},
\end{equation}
where $\Theta_{q}$ is the quantity conjugate to $N_{q}$. The Hawking temperature, angular velocity, and pressure $\Theta_{q}$ can be written as
\begin{eqnarray}\label{tomth}
T_+=\left(\frac{\partial M}{\partial S}\right)_{J, N_{q}}, \quad \Omega_+=\left(\frac{\partial M}{\partial J}\right)_{S,N_{q}}, \quad \Theta_q= \left(\frac{\partial M}{\partial N_{q}}\right)_{S,J}.
\end{eqnarray}
The first law of thermodynamics has the following form
\begin{equation}
d M = T_+ dS + \Omega_+ dJ + \Theta_q d N_q.
\end{equation}
Now, we focus on the thermodynamic stability of the solution (\ref{fr}). The black string is thermodynamically stable/unstable according to the heat capacity is positive/negative. The expression for the heat capacity of a thermodynamic system is written as
\begin{eqnarray}
C_+&=&\left(\frac{dm}{dT}\right)_{(r=r_{+})}.  
\end{eqnarray}
\begin{figure}[t]
\includegraphics[scale=0.5]{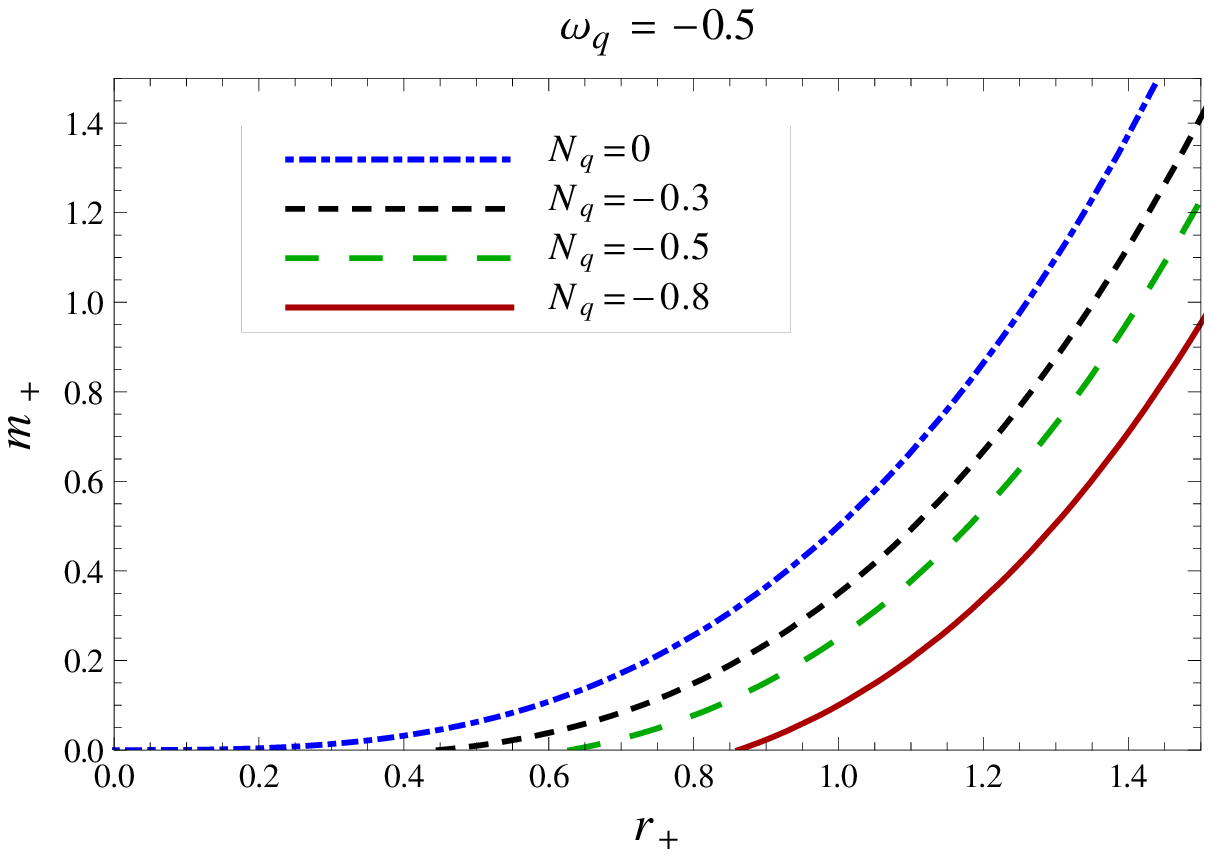}
\includegraphics[scale=0.5]{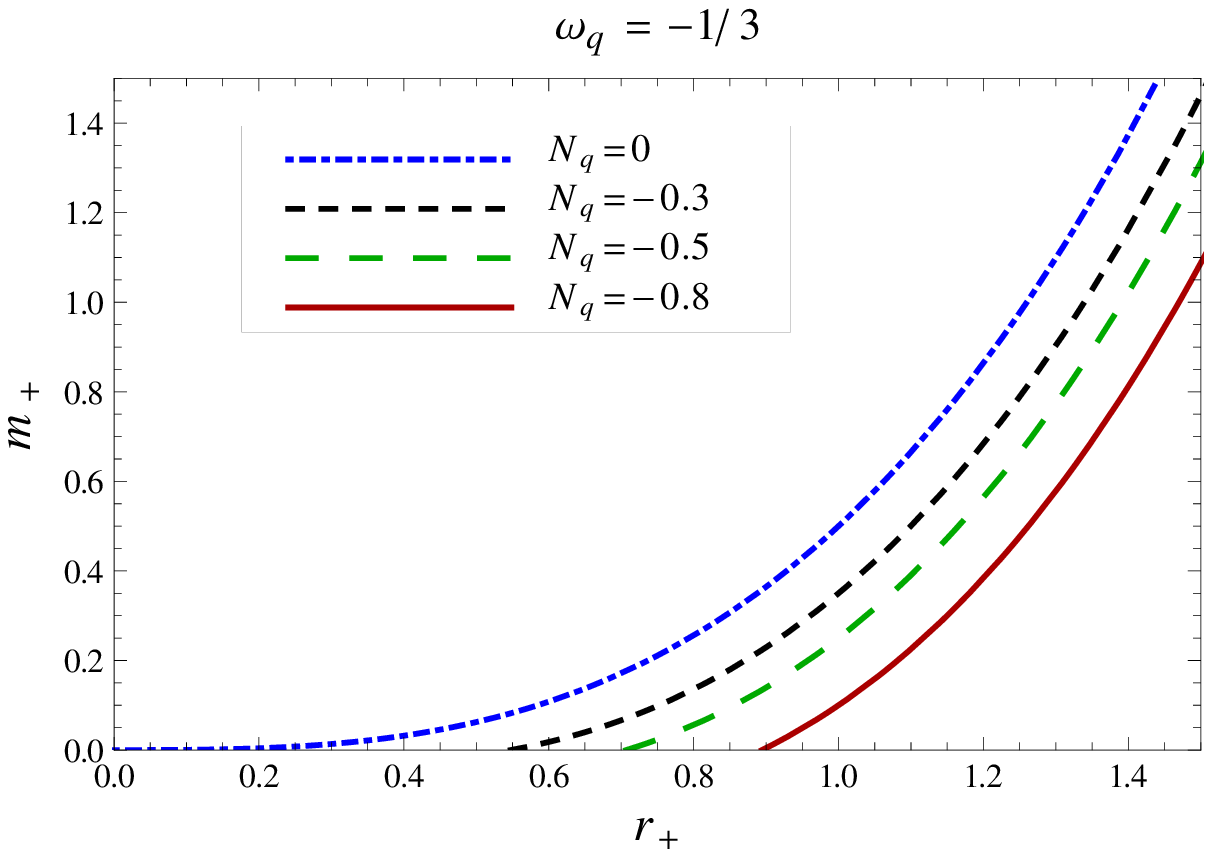}
 \caption{Plots showing the behaviour of mass function $m_+$ with horizon radius $r_+$ for rotating black string sorrounded by quintessence for various values of $N_q$ in units of $l^{3w_q+1}$. Here both the horizon radius $r_+$ and mass $m_+$ are written in units of $l$. \label{fig3}}
\end{figure}
The mass of the black hole in the quintessence background can be calculated from Eq.~(\ref{fr}) and given in terms of event horizon radius as 
\begin{eqnarray}
m_+= \frac{r_+}{2}\left(\frac{r_+^2}{l^2}+\frac{N_q}{r_+^{3 w_q+1}}\right).
\end{eqnarray}
The behavior of mass $m$ is shown in Fig.~\ref{fig3}. One can see that mass of the black string is ever increasing function of horizon radius $r_+$.
\begin{figure}[t]
\includegraphics[scale=0.5]{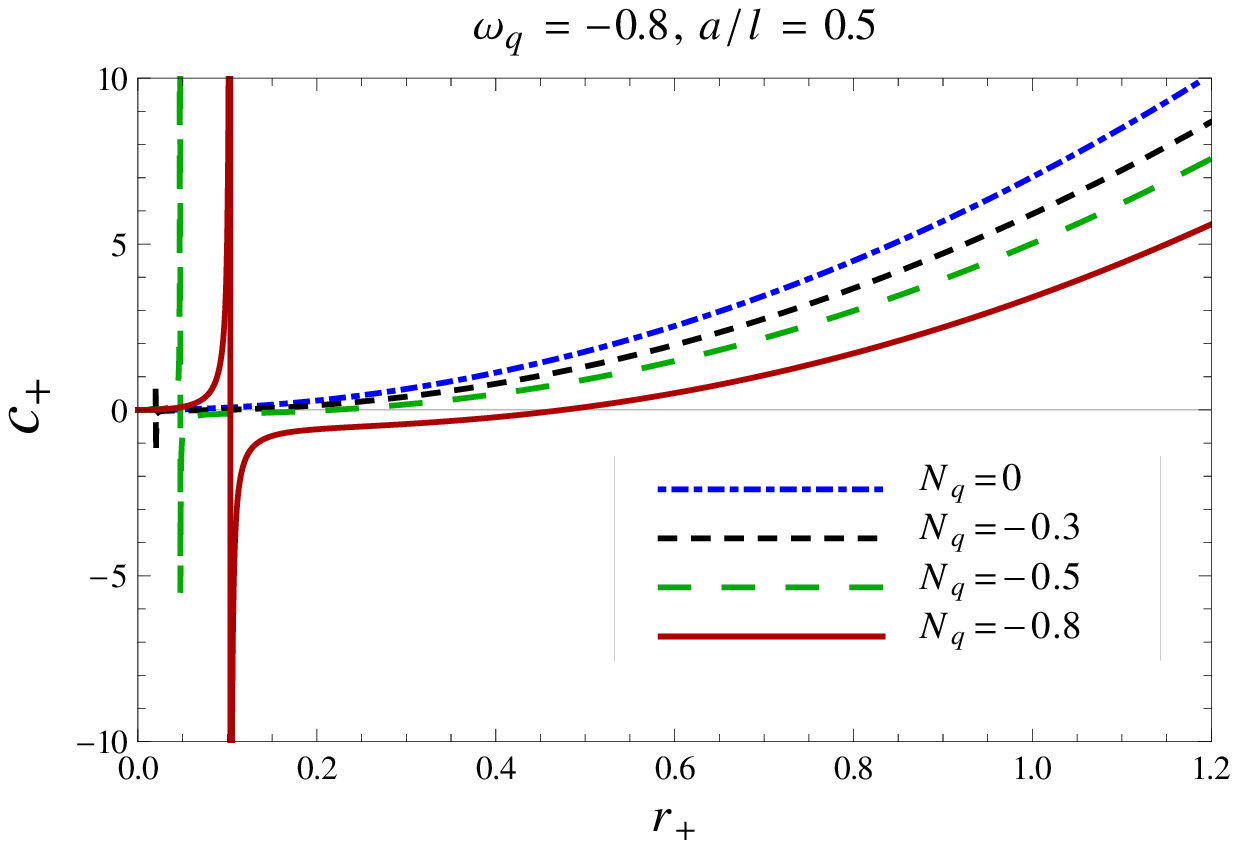}
\includegraphics[scale=0.5]{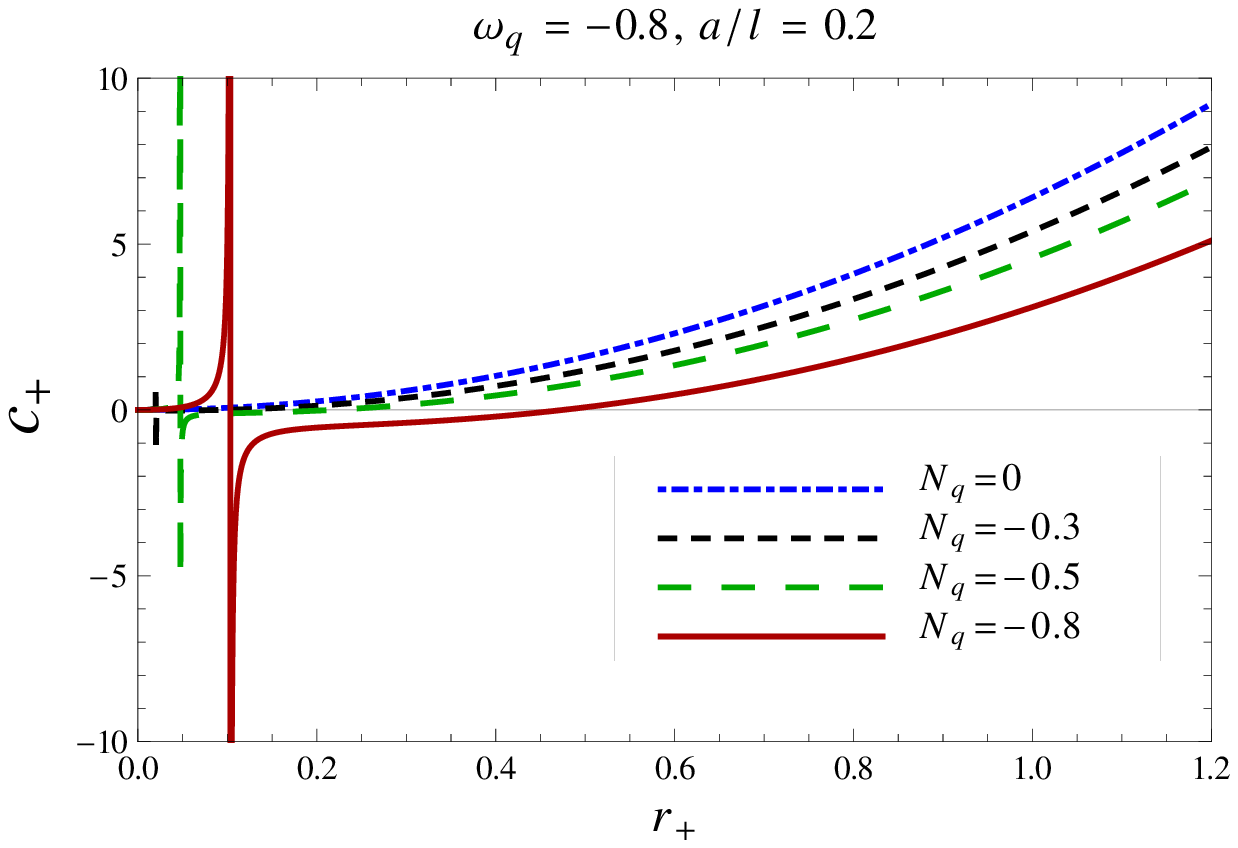}
 \caption{Plots showing the behaviour of heat capicity $C_+$ with horizon radius $r_+$ for rotating black string sorrounded by quintessence for various values of $N_q$ in units of $l^{3w_q+1}$. Here the horizon radius $r_+$ is written in units of $l$ and $C_+$ in units of $l^2$. \label{fig4}}
\end{figure}
The heat capacity for the metric~(\ref{fr}) is 
\begin{eqnarray}
C_{+} &=&\frac{ 2 \pi r_+^2 \Xi \left(\frac{r_+^2}{l^2}-\frac{N_q w_q}{r_+^{3 w_q+1}}\right)}{\left(\frac{r_+^2}{l^2}+\frac{\left(3 w_q+2\right)N_q w_q}{r_+^{3 w_q+1}}\right)}.
\end{eqnarray}
One can see the behavior of the heat capacity from the Fig.~\ref{fig4} for various values of quintessence parameter $N_q$, state parameter $w_q$, and rotation parameter $a$. It turns out that for a given set of parameters, there exist a critical radius $r^c$, when $C_+ \rightarrow \infty$. The critical radius $r^c$ can be written as
\begin{eqnarray}
r^c=\left[-(3w_q+2)w_q N_q l^2\right]^{\frac{1}{3w_q+3}}.
\end{eqnarray}
The critical radius has positive values when $(3w_q+2)w_q>0$, which leads to two different possibilities such that
$$\mbox{i}.\;(3w_q+2)<0 \;\text{and}\;w_q<0,$$ 
$$\mbox{ii}.\;(3w_q+2)>0 \;\text{and}\;w_q>0.$$
The condition ($i$) implies $w_q<-2/3$, while the condition ($ii$) implies $w_q>-2/3$ or $w_q>0$, which is unphysical. Thus we discard the condition $(ii)$, and therefore the range of $w_q$ lies between  $-1<w_q<-2/3$.
Thus, the heat capacity $C_+<0(>0)$, respectively, when $r<r^c(>r^c)$ and hence the black string is always thermodynamically stable when $r>r^c$. The heat capacity for $N_q=0$ turns out to be
\begin{eqnarray}
C_{+} &=&2 \pi\Xi r_+^2,
\end{eqnarray}
which is a parabola and thus the black string is thermodynamically stable with positive heat capacity.

\subsection{Cloud of strings}\label{cloudstring}
\begin{figure}[t]
\includegraphics[scale=0.5]{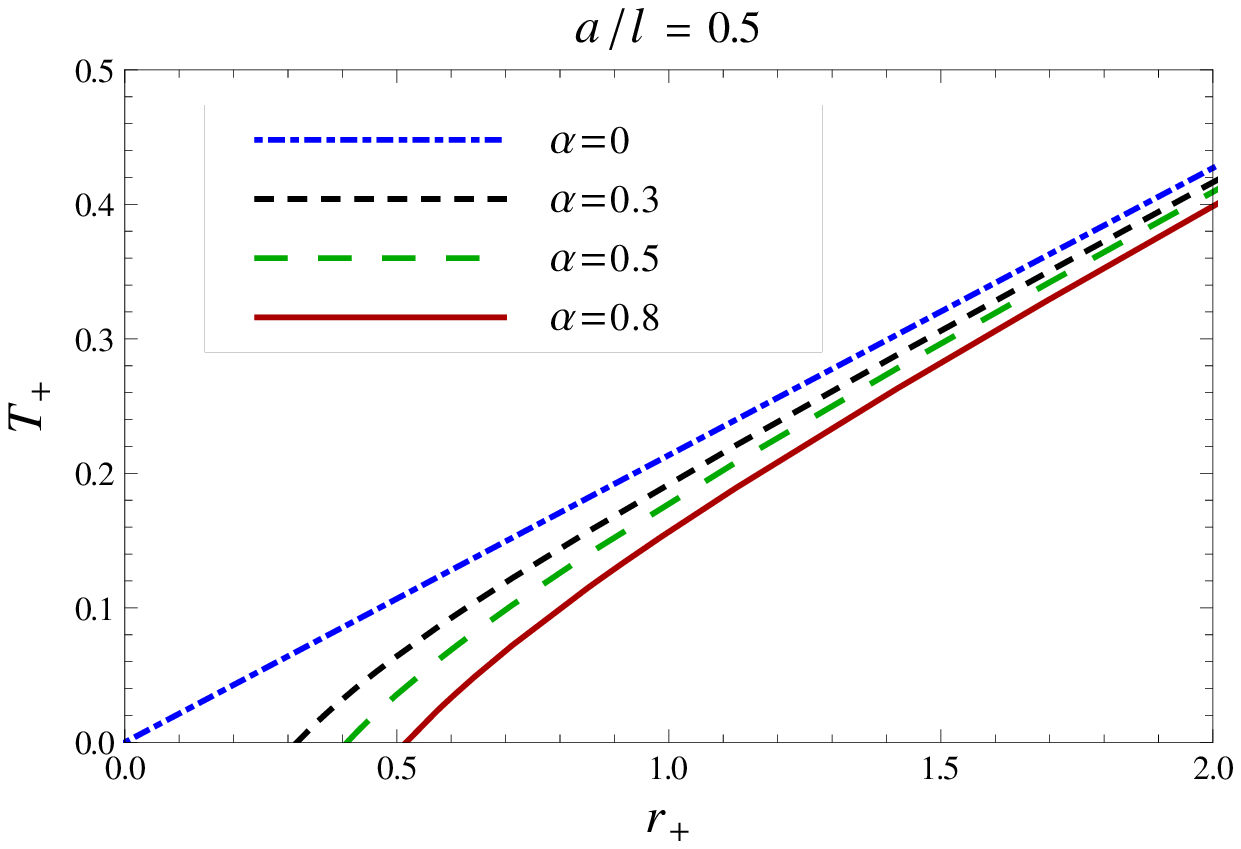}
\includegraphics[scale=0.5]{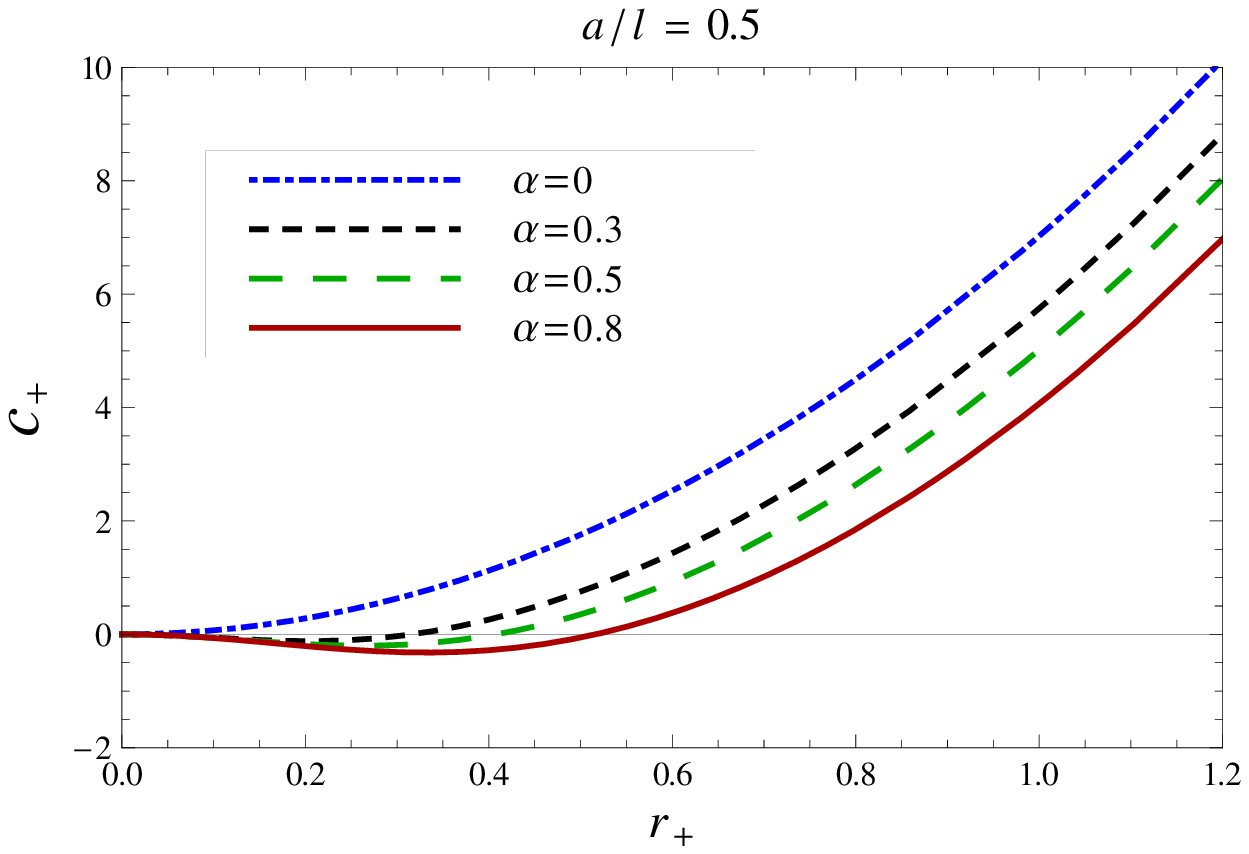}
 \caption{Plots showing the behaviour of temperature $T_+$ ($left$) and heat capacity $C_+$ ($right$) for string cloud background for various values of $\alpha$. Here the horizon radius $r_+$, the temperature $T_+$, and the heat capacity $C_+$ are written, respectively, in units of $l$, $1/l$, and $l^2$. \label{fig7}}
\end{figure}

Here, we consider a special case of black string solution (\ref{rotBSQ}), when $w_q=-1/3$. In this surrounding matter is a cloud of strings, which is defined as an arrangement of one-dimensional strings and could be effective in the strong gravity regions such as black holes. The strings are assumed to be fundamental objects in nature which encourages us to analyze its consequences on various gravitational theories. The black string solutions in the string theory are the one-dimensional extended objects surrounded by the event horizon. Letelier obtained the first solution of the black string in the background of a cloud of strings \cite{Letelier:1979ej}. After that many other solutions have been obtained \cite{Herscovich:2010vr,Ghosh:2014dqa,Ghosh:2014pga,Lee:2014dha,Toledo:2018pfy}, and the thermodynamics have been discussed \cite{Toledo:2018hav,Costa:2018opi}.
In this energy-momentum tensor reads
\begin{equation}
T^t{_t}=T^r{_r}=\frac{\alpha}{r^2},\quad
T^\phi{_\phi}=T^z{_z}=0,
\end{equation}
where $N_q=-\alpha$. The metric~(\ref{rotBSQ}) simplifies to
\begin{eqnarray}
ds^{2}=-\left(\frac{r^2}{l^2}-\frac{2m}{r}-\alpha\right) \left(\Xi dt-a d\phi\right)^2+\frac{dr^2}{\left(\frac{r^2}{l^2}-\frac{2m}{r}-\alpha\right)} +\frac{r^2}{l^4}\left(adt-\Xi l^2d\phi\right)^2 +\frac{r^2}{l^2} dz^2. 
\end{eqnarray}
The quasilocal mass and the angular momentum using the $(3+1)$ formalism have been calculated for the quintessence background. Now we can write these conserved quantities for the cloud of string background 
\begin{eqnarray}
&& M=\frac{r_+}{32\pi l}\left(3\Xi^2-1\right)\left(\frac{r_+^2}{l^2}-\alpha\right),\nonumber\\
&& J =\frac{3\Xi a r_+}{32\pi l}\left(\frac{r_+^2}{l^2}-\alpha\right).
\end{eqnarray}
The entropy, the Hawking temperature, and the heat capacity for the cloud of strings background are given by
\begin{eqnarray}
&& S_+=\frac{\pi\Xi r_{+}^2}{4 l}, \\
&& T_+=  \frac{1}{4 \pi r_+ \Xi}\left(3 \frac{r_+^2}{l^2}-\alpha\right),\\
&& C_+=\frac{2 \pi r_+^2 \Xi \left(3 \frac{r_+^2}{l^2}-\alpha\right)}{\left(3 \frac{r_+^2}{l^2}+\alpha\right)}.
\end{eqnarray}
The temperature of the black string in the cloud of strings background is positive, zero, and negative, respectively, when $r_+^2>\alpha l^2/3$, $r_+^2=\alpha l^2/3$, and $r_+^2<\alpha l^2/3$.
The heat capacity can be positive or negative depending on the values of the horizon radius. If $r_+^2>\alpha l^2/3$, then it is positive and the solution is thermodynamically stable. On the other hand when $r_+^2<\alpha l^2/3$, the black string solution is thermodynamically unstable and it decays through the Hawking radiation. The Hawking temperature and the heat capacity for the rotating black string in the cloud of strings background have been depicted in Fig.~\ref{fig7}. The heat capacity is negative for small values of $r$ therefore indicates that the black string with a smaller radius is thermodynamically unstable. For large values, it is always positive. Therefore, the black string with a larger horizon radius do not evaporate through Hawking radiation and becomes thermodynamically stable. However, when the horizon radius $r_+^2=\alpha l^2/3$, the heat capacity changes its sign from negative to positive values.
\section{Conclusions}
\label{conclusion}
The first exact spherically symmetric black hole solution of the vacuum Einstein equations was obtained by Schwarzschild \cite{schw}, and five decades later, in 1963, the solution of a rotating black hole was discovered by Kerr \cite{Kerr:1963ud}. If one considers cylindrical symmetric Schwarzschild like solutions in AdS, it describes a black string \cite{Lemos:1994fn} and similarly, the Kerr spacetime turns into a rotating black string \cite{Lemos:1998iy}. We obtained a static cylindrically symmetric black string solution with quintessence matter, which is asymptotically anti-de Sitter and investigated their thermodynamic properties. Our rotating black string solution encompasses the Lemos's black string \cite{Lemos:1994xp} which can be recovered in the absence of quintessence matter ($ N_q=0 $). Further, the charged black string \cite{Lemos:1995cm} is a special in the limit $w_q=1/3$ and black string surrounded by clouds of string matter is obtained as a special case for $w_q=-1/3$. Interestingly, even in the absence of the cosmological constant ($1/l^2=0$), one can obtain a black string solution when quintessence state parameter $w_q=-1$. It turns out that, unlike Kerr black hole, the rotating black string (\ref{rotBSQ}) admits just one horizon except for the charge black string. Thus, the extremal black string with degenerate horizons is possible only for the charged black string.  Indeed, we obtained mass and angular momenta of the black string surrounded by quintessence background via the counterterm method. The thermodynamic quantities, the entropy, the Hawking temperature, and the angular velocity of the rotating black string have been obtained. The entropy of a black string is not affected by the background quintessence matter, which still obeys the area law. The thermodynamical stability is performed using the heat capacity of the rotating black string. In the special case when $w_q=-1/3$, the rotating black string is thermodynamically stable for $r_+^2>\alpha l^2/3$ with positive capacity, unstable when $r_+^2<\alpha l^2/3$ with negative heat capacity. Finally, we concluded that the conserved and thermodynamic quantities obey the first law of thermodynamics.

\acknowledgements
S.G.G. thanks DST INDO-South Africa (INDO-SA) bilateral project DST/INT/South Africa/P-06/2016. MSA research is supported by the ISIRD grant 9-252/2016/IITRPR/708.

\end{document}